\documentclass{INTERSPEECH2023}

\interspeechcameraready 

\usepackage{arydshln}
\usepackage{makecell}

\title{Speech-to-Text Adapter and Speech-to-Entity Retriever \\ Augmented LLMs for Speech Understanding}
\name{Mingqiu Wang, Izhak Shafran, Hagen Soltau, Wei Han, Yuan Cao, Dian Yu, Laurent El Shafey}
\address{Google DeepMind}
\email{mingqiuwang,izhak,soltau,weihan,yuancao,dianyu,shafey@google.com}

\begin{document}

\maketitle

\begin{abstract}

Large Language Models (LLMs) have been applied in the speech domain, often incurring a performance drop due to misaligned between speech and language representations. To bridge this gap, we propose a joint speech and language model (SLM) using a Speech2Text adapter, which maps speech into text token embedding space without speech information loss. Additionally, using a CTC-based blank-filtering, we can reduce the speech sequence length to that of text. In speech MultiWoz dataset (DSTC11 challenge), SLM largely improves the dialog state tracking (DST) performance  (24.7\% to 28.4\% accuracy). Further to address errors on rare entities, we augment SLM with a Speech2Entity retriever, which uses speech to retrieve relevant entities, and then adds them to the original SLM input as a prefix. With this retrieval-augmented SLM (ReSLM), the DST performance jumps to 34.6\% accuracy. Moreover, augmenting the ASR task with the dialog understanding task improves the ASR performance from 9.4\% to 8.5\% WER.

\end{abstract}

\section{Introduction}
There has been considerable interest in extending the capability of the large language models (LLMs) from text to other modalities including speech. One thread of work attempts to map speech and text to the same latent representations ~\cite{bapna-etal-2021-slam, thomas2020, chen22r_interspeech}. A {\em shared encoder} is employed for both speech and text, in one case with an explicit loss term promoting the same embedding space~\cite{chen22r_interspeech} and in other without the explicit term~\cite{bapna-etal-2021-slam}.

In most practical spoken language systems, speech input is recognized using an automatic speech recognition (ASR) and the recognized transcripts are fed into a back-end NLU system, where the back-ends are increasing powered by LLMs~\cite{zhao2022description}. This cascaded approach does not offer an opportunity to correct potential ASR misrecognitions. Besides, both the LLMs and the ASR systems have a common weakness in processing entities that are not well-represented in their training data.

In this paper, we examine these challenges in the context of a speech understanding system using the DSTC-11 dialog tracking task~\cite{soltau2022speech}. The task is a based on the popular MultiWoz, a fully-labeled collection of human-human conversations spanning multiple domains and topics such as train and hotel reservations~\cite{multiwoz21}. In this particular challenge, the written user responses were replaced with spoken version collected from crowd-sourced workers. The model is expected to infer the dialog states corresponding to the current user utterance and given dialog context. The context could be the acoustic or the recognized version of the dialog history along with the previously inferred states. This task is particularly interesting because of high occurrence of rare entities such as restaurants, tourist attractions, cities and train stations.

\begin{figure}[h]
    \centering
    \includegraphics[width=0.5\textwidth]{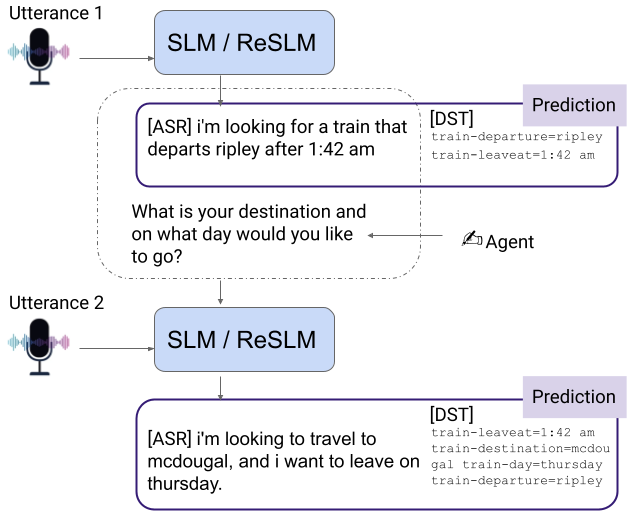}
    \caption{Direct speech to dialog state prediction in multi-turn dialogs.
    Given speech of the current user turn $i$ and a text transcript of the dialog history, the SLM / ReSLM models generate a single output sequence for both the corresponding transcript [ASR] and dialog state [DST]. The ASR transcript predicted from turn $i$ is used as history turn $i+1$ in an auto-regressive manner.}
    \label{fig:auto-regressive}
    \vspace*{-.25in}
\end{figure}

The key contributions of this paper are: 
\begin{enumerate}
    \itemsep0em 
    \item We propose a Speech2Text adapter which maps speech encodings into text token embeddings with seemingly minimal loss of information.
    \item We propose a joint speech and language model (SLM) with both speech and text modalities in input. With the adapter, SLM can leverage the rich knowledge encoded in LLMs (pretrained T5 XXL in our case) directly for speech tasks.
    \item We introduce a Speech2Entity retriever to handle rare task-specific entities, which uses speech to retrieve entities.
    \item We propose a novel retrieval-augmented speech language model (ReSLM) which augments SLM with the retriever. We show that ReSLM achieves strong performance in ASR and speech understanding tasks.
\end{enumerate}
Unlike cascaded systems, both SLM and ReSLM operate directly on speech input and as such are not stuck with misrecognized words from the first stage ASR.
We demonstrate the benefits of the different components of the model using the DSTC11 challenge task. While this work reports results on a dialog tracking task, the model is applicable more widely for speech understanding tasks. 

\section{Related work}

A closely related line of work injects text inputs into speech models~\cite{rosenberg_etal_2019_speech, bapna-etal-2021-slam, chen22r_interspeech, chen-etal-2022-tts4pretrain, thomas2020} and align the learned representation of text and speech to the same space. This is done by TTS or more recently {\em up-sampled} text and minimizing an L2 loss of aligned speech and text frames. This is in contrast to our work, where we do the opposite and reduce the frame rate of the audio sequence to bring it closer to text. This is done via a CTC model~\cite{graves_ctc} where we use the predictions to filter out blank frames. This results in a highly compressed speech encodings that preserve semantic information and makes down-stream NLU tasks much easier.  There also other use cases of filtering CTC blank frames. For example, the work in~\cite{ctcrnntblank2022} used it to speed up training of RNN-T~\cite{Graves2012} models.

The compression of speech signal when combinining speech and text modalitiess has analogies in tasks where vision and text modalities are combined. For example, a Perceiver~\cite{Perceiver} architecture is used to compress images before interleaved with text tokens~\cite{flamingo}. However, the cross-attentions between the Perceiver outputs and the frozen LM layers makes the model substantially different from a standard LM and hence their model cannot share a standard LM at serving time, unlike our work.

In an alternative approach, the speech input is tokenized and then fed into the LLMs~\cite{borsos2022audiolm,wav2seq-2022}. These approaches suffer from the same issue as cascaded systems where LLMs cannot utilize acoustic encodings to correct errors in tokenization.


Retrieval-augmented language models have demonstrated superior performance on various natural language tasks~\cite{khandelwal-etal-2021-nearest, borgeaud-etal-2021-improving, izacard-etal-2022-atlas}, especially the knowledge-intensive ones~\cite{guu-etal-2020-retrieval, lewis-etal-2020-retrieval, izacard-grave-2021-distilling}. In particular, retrieval-augmentation disentangles encoding domain-specific knowledge from training parameters, thereby being a desirable property for task-oriented dialog where integrating domain schema is a critical requirement~\cite{wu-etal-2019-transferable, zhou-small-2019-multi}. Furthermore, in dialog understanding, unseen domains and tasks may demand adaptation to a new set of schema~\cite{rastogi-etal-2020-towards}. To deal with these challenges, previous work propose to either retrieve similar training examples~\cite{pasupat-etal-2021-controllable, gupta-etal-2022-show} or corresponding intents and slots for dialog state tracking~\cite{yu-etal-2022-knowledge}. 
 
Inspired by these work, we extend retrieval-augmented methods to speech understanding. As mentioned before, unlike written domain, speech understanding poses an additional challenge that rare entities are not easily recognizable~\cite{soltau2022speech}. Therefore we introduce an audio retrieval method to alleviate these difficulties and achieve better performance on end-to-end speech dialog understanding.

\section{Model}

\subsection{Joint speech and language model (SLM)}

\begin{figure}[h]
    \vspace{-.1in}
    \centering
    \includegraphics[width=0.5\textwidth]{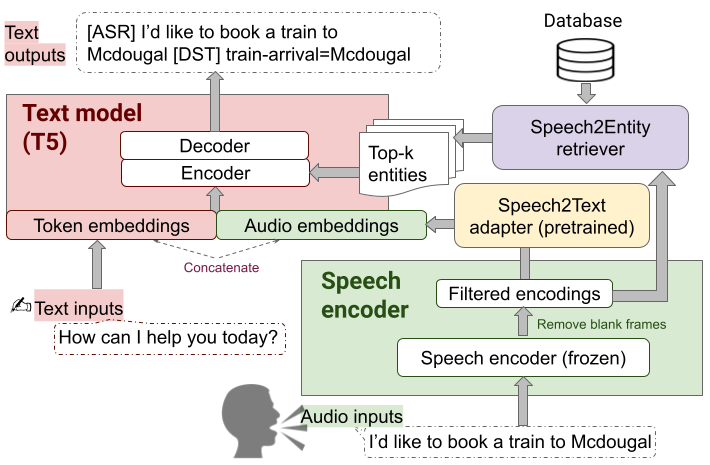}
    \caption{Model architecture for ReSLM, and SLM (without the Speech2Entity retriever component). The SLM and ReSLM models take both speech and text as inputs. The speech frame sequence is shortened by CTC-based blank-filtering, transformed by Speech2Text adapter, then concatenated with text embeddings before being fed into a T5 encoder-decoder model. In ReSLM, a few entities are selected using the speech input by Speech2Entity retriever and prepended to the text input.}
    \label{fig:reslm}
    \vspace{-.2in}
\end{figure}

The speech understanding task requires a model that can simultaneously recognize speech inputs and understand the semantic meaning of the corresponding language. In previous research, large pre-trained language models such as BERT, T5, and GPT have demonstrated impressive capabilities for understanding semantics in NLU tasks~\cite{devlin2018bert, raffel2020exploring, radford2018improving}. Leveraging this capability, we combine a speech encoder with a T5 model for this speech understanding task as shown in Figure~\ref{fig:reslm}.

The speech encoder is based on a CTC model trained separately, described further in Section~\ref{sec:expts}. We only utilize non-blank predicted frames of the CTC model.
This CTC-based blank-filtering approach has two advantages: First, only semantic relevant information of the audio signal is being 'forwarded' to the down-stream task, making fine-tuning for NLU much easier. Secondly, the effective sequence length of the encoded speech sequence is reduced by approximately 4x. This helps with joint modeling of speech and text sequences, where otherwise the audio sequence is much larger than the text sequence and makes processing much harder. Note, this is in contrast to the opposite approach employed in other works where text was upsampled to match the acoustic frame rate~\cite{rosenberg_etal_2019_speech} which cannot take advantage of pre-trained LLMs. 

\subsection{Speech2Text Adapter}
\label{adapter}
The Speech2Text adapter consists of a few self-attention layers to map the CTC-filtered speech encodings to the text token embeddings of the pre-trained LLMs. The resulting outputs are concatenated with the text embeddings and then fed into the pre-trained LLMs. Note that for the adapter to be effective, it is crucial that it undergoes pre-training to ensure a successful mapping to the text embedding space. This can be done by simply training SLM with any ASR task, where the input is the speech and the prediction is the corresponding transcript. The text input part of SLM is unused while training the adapter. It's worth noting that both the speech and language model weights are frozen during this pre-training process.

Therefore, our Speech2Text adapter refers to two folds of meanings: 1) a few self-attention layers between speech encoder and language model; 2) pretraining with both speech and language models frozen.

\subsection{Speech2Entity Retriever} \label{sec:audio_retriever}
The main task of the retriever is to extract a subset of entities from a list of given task-specific entities that are relevant for the current speech input. We adopt a dual encoder architecture for the retriever whose keys are acoustic encodings of the speech input and the values are the entities mentioned in the input~\cite{ni2021large}. The model is trained using entities mentioned in the reference transcript of the input speech. The keys and values (candidate entities) are encoded separately and cosine distance between their representations is used to measure similarity. The in-batch negatives are used as negative targets to optimize the contrastive loss. In our case, we use the multimodal SLM encoder since it can encode both audio and text. During inference, we compute the nearest neighbors efficiently using cosine distance with the SCAM library and retrieve the top-K candidates~\cite{guo2020accelerating}.

\subsection{Retrieval-augmented SLM model (ReSLM)} \label{sec:reslm}

In the retriever-based SLM, we integrate the top-K candidates from the audio retriever into the previously described SLM. Specifically, with acoustic encodings of the current speech input as queries we retieve the top-K entities from the large pool of task-specific entities. The retrieved entities are pre-prended to the original text inputs before being fed into the encoder.

\section{Experiments and results}
\label{sec:results}

\subsection{Evaluation Task}
The DSTC11 Challenge Task is based on MutliWoz 2.1 and has the same distribution in terms of dialogs and turns~\cite{soltau2022speech}. The main difference is that the written user responses are replaced with spoken versions. The responses were generated using TTS in the training set and by human voices from crowd-sourced workers in the test set. Additionally, previously researchers had discovered that the slot values in the training and test sets had substantial overlap, which led to misleading and overly optimistic performance reports. To alleviate this issue, the organizers of the challenge modified the test set by replacing the slot values (city, restuarant names, times, etc). As such, the performance of systems on DSTC11 test set are expected to be lower than the written version. 

The main focus of the task was dialog state tracking where the performance was measured using Joint Goal Accuracy (JGA) and Slot Error Rate (SER). For details, see~\cite{soltau2022speech}. Additionally, we also measure word error rate (WER) of the recognized input speech for ablation experiments to tease apart the impact of misrecognitions.

\subsection{Experiment Setup} \label{sec:expts}
The speech encoder is derived from a CTC~\cite{graves_ctc} model trained on the PeopleSpeech public corpus~\cite{PSdata} of approx. 32,000 hours of speech data. The encoder consists of 16 Transformer layers, altogether a 220m parameter model. The model's input frame rate is 25ms and produces outputs every 75ms obtained via a down-sampling layer sandwiched between the transformer layers. We use the activations (1024-dim) from the last transformer layer as speech encodings. Additionally, we remove {\em blank} frames (e.g. frames where the highest scoring token is blank). The model emits a non-blank frame on average every $305ms$ and each word is encoded on average with $1.48$ frames. Filtering CTC blank frames results in a very strong compression of the speech signal and makes down-stream NLU tasks substantially easier while preserving the semantic information.

We reused the previously trained unimodal checkpoints: specifically the T5 XXL checkpoints for the text encoder-decoder, and the CTC speech encoder checkpoints for the speech encoder. Throughout the training process, we maintained the speech encoder in a frozen state for all experiments and exclusively trained the text encoder-decoder along with the Speech2Text adaptation layer. We also show ablation studies of only partially finetuning T5 models.

\subsection{Auto-Regressive Inference} 
Dialogs have multiple turns and the dialog state values are inferred turn-by-turn auto-regressively. The task of dialog state tracking requires predicting all dialog states up to the current turn $i$, therefore the entire dialog history is required as input.

As shown in Figure \ref{fig:auto-regressive}, we feed the speech of turn $i$ as speech input and the dialog history from turn $1$ to $i-1$ as text input. The dialog history can be long and is best represented in the text form, not speech. For this reason, we trained the SLM model to simultaneously recognize the words spoken in turn $i$ along with the dialog states in one output string. The transcript from each turn is incrementally collated to create the dialog history for subsequent turns.

During the training process, the input consists of speech of the current turn and the dialog history based on the ASR transcripts from the previously described CTC model. The loss is computed with target consisting of the reference transcript of the current turn and the associated reference dialog state values.

\subsection{Speech2Entity Retriever Results}
The Speech2Entity retriever, describe in Section~\ref{sec:audio_retriever}, was trained on three categories of the entities: hotel names, restaurant names, and  city names~\cite{soltau2022speech}. The retriever was trained on a pool of 2.5k entities and a separate pool of 14k entities were used for evaluation. 

In principle, the two-tower retriever model can utilize any speech and text encoders. In our experiments, we use the SLM speech and text encoders both query and candidates. The checkpoints from previously trained SLM was used to initialize the two-tower encoders before training the retriever.

The performance of the Speech2Entity retriever is shown in Table~\ref{retriver}. The subset of retrieved entities were selected using a distance threshold of -0.78, which resulted in top-10 entities per utterance. This threshold was chosen to balance recall and precision, with a focus on optimizing recall so that the resulting ReSLM model could access entities with the highest possible coverage. As a consequence of this emphasis on recall, precision was sacrificed. However, we anticipated that the ReSLM model would learn to discard incorrectly retrieved entities. Clearly, the retriever can be improved further and this is mostly a demonstration of the proof-of-the-concept and in spite of the poor precision we obtain substantial gains in ReSLM as described later.

\begin{table}[h]
  \centering
  \begin{tabular}{ c c c c}
    \toprule
   \multicolumn{2}{c}{\textbf{Recall (\%)}} &
    \multicolumn{2}{c}{\textbf{Precision (\%)}} 
    \\
    \midrule
    R@1  & 40.2  & P@1 & 13.3 \\
    R@3  & 51.9  & P@3 & 6.7 \\
    R@5  & 57.0  & P@5 & 5.0 \\
    \underline{R@10}  & \underline{62.2}  & \underline{P@10}  & \underline{3.6}           \\
    R@20  & 66.5 & P@20  & 2.8            \\
    R@100 & 70.4 & P@100 & 2.0            \\
    \bottomrule
  \end{tabular}
 \caption{Performance of the Speech2Entity retriever. Top-k recall and precision filtered by -0.78 similarity threshold.}
\label{retriver}
\vspace*{-.3in}
\end{table}

\subsection{Dialog State Tracking Results}

The results on dialog state tracking are reported in the Table~\ref{jga} where the left half corresponds to the SLM model and the right to the ReSLM. In the upper half of the table, the adapter layers were trained from scratch and in lower half of the table, the Speech2Text adapter was trained with ASR task. The different rows shows the impact of training different groups of parameters including the embedding layer, the encoder and the decoder of the T5 model.

The results show that the Speech2Text adapter improves performance for both SLM and ReSLM, with gains ranging from 3-5 JGA. Interestingly, when Speech2Text adapter is employed just training the encoder and/or embedding gives the best result (29.8\% and 35.1\% JGA), suggesting that adapter is effective is bringing the speech modality close to the text modality. On top of the gains from Speech2Text adapter, the Speech2Text retriever gives a further boost of ~5\% JGA in all conditions. 

\begin{table}[h]
\vspace{-.1in}
\centering
\begin{tabular}{lcccc}
\toprule
 $\%$ & \textbf{JGA} $\uparrow$ & \textbf{WER} $\downarrow$ & \textbf{JGA} $\uparrow$ & \textbf{WER} $\downarrow$ \\ 
\midrule
Cascaded \cite{soltau2022speech} & 31.8 & 13.0 & & \\
\midrule
\midrule
\textbf{Trainable params} & \multicolumn{2}{c}{SLM} & \multicolumn{2}{c}{ReSLM}\\
\midrule
\multicolumn{5}{c}{without Speech2Text Adapter} \\
\hdashline
Whole T5        & 24.7 & 11.5 & 31.3 & 9.5 \\
T5 encoder+emb  & 27.3 & 10.1 & 32.0 & 8.9 \\ 
T5 encoder only & 27.1 & 11.6 & 31.6 & 9.3 \\ 
\midrule
\multicolumn{5}{c}{with Speech2Text Adapter} \\
\hdashline
Whole T5        & 28.4 & 9.2 & 34.6 & 8.5  \\
T5 encoder+emb  & 29.5 & 9.2 & 35.1 & 8.5  \\ 
T5 encoder only & 29.8 & 9.2 & 34.5 & 8.6  \\ 
\bottomrule
\end{tabular}
\caption{Dialog state tracking performance evaluated using joint goal accuracy (JGA). We compare model performances with and without pretrained (see section \ref{adapter}) Speech2Text adapters, with and without retrieved entities. Note that the SLM / ReSLM models predict both speech recognized transcript and dialog state in the same output sequence. So we can also report word error rate (WER) here. All numbers are on test set.
}
\label{jga}
\vspace{-.3in}
\end{table}

Zooming into the improvements for categories of dialog state variables, using Slot Error Rate (SER), show that the following categories of dialog state variables benefited from the Speech2Text retriever: hotel names (26\% gain), train destination station (35\% gain), and and restaurant name (14\% gain). This is in spite of poor precision of the Speech2Text retriever (see section below), which makes this a remarkable gain.

\subsection{ASR Results} 
Since we trained the model using multi-task objective to include ASR, we can evaluate the performance of the model on the recognition task. There are two clear trends, the Speech2Text adapter improves the ASR performance across all conditions for SLM. It also compares favorably with a general purpose baseline RNN-T ASR model ($13.0\%$ WER)~\cite{soltau2022speech}. When Speech2Entity retriever is also used the gain is further boosted in all cases, mirroring the results in dialog state tracking.

One useful ablation study would be to understand the ASR gains without the DST loss. We tested this by feeding speech input and training the model for ASR task alone and report the results in Table~\ref{asr}. The ASR performance matches the in-domain ASR system ($10.4\%$ vs $10.7\%$). Since we were able to achieve this performance while keeping the LLM frozen, we hypothesize that the Speech2Text adapter is able to map from the acoustic encoding space to the textual encoding space. Interestingly, while the Speech2Text retriever does not bring additional gains when trained on ASR task alone, it brings gains when trained with DST task (Table~\ref{jga}). This can be attributed to the fact that the DST loss places additional focus on improving the recognition of entities and semantics of their context.

\begin{table}[h]
\vspace{-.1in}
\centering
\begin{tabular}{lcc}
\toprule
 \% &  \multicolumn{2}{c}{\textbf{WER} $\downarrow$}\\ 
\midrule
RNN-T \cite{soltau2022speech} & \multicolumn{2}{c}{$13.0$} \\
RNN-T in-domain finetuned \cite{soltau2022speech} & \multicolumn{2}{c}{$10.4$} \\
Adapter only & \multicolumn{2}{c}{$10.7$}
\\
\midrule
\midrule
\textbf{Trainable params} & SLM & ReSLM\\
\midrule
\multicolumn{3}{c}{without Speech2Text Adapter} \\
\hdashline
Whole T5        & 12.0 & 11.0 \\
T5 encoder+emb  & 11.2 & 11.2 \\ 
T5 encoder only & 11.3 & 10.1 \\ 
\midrule
\multicolumn{3}{c}{with Speech2Text Adapter} \\
\hdashline
Whole T5        & 9.7 & 9.4 \\
T5 encoder+emb  & 9.4 & 9.5 \\ 
T5 encoder only & 9.5 & 9.4 \\ 
\bottomrule
\end{tabular}
\caption{Speech recognition performance. We compare model performances with and without pretrained (see section \ref{adapter}) Speech2Text adapters, with and without retrieved entities. The WER values here were calculated from an ASR-only setup (different from joint DST-ASR setup in Table \ref{jga}).}
\label{asr}
\vspace{-.2in}
\end{table}

\begin{table}[h]
\vspace{-.1in}
\centering
\begin{tabular}{lc}
\toprule
 $\%$&  \textbf{WER} $\downarrow$\\ 
\midrule
RNN-T \cite{soltau2022speech} & $13.0$ \\
RNN-T in-domain finetuned \cite{soltau2022speech} & $10.4$ \\
Adapter only & $10.7$
\\
\midrule
\midrule
\textbf{Trainable params} & SLM \\
\midrule
\multicolumn{2}{c}{without Speech2Text Adapter} \\
\hdashline
Whole T5        & 12.0 \\
T5 encoder+emb  & 11.2 \\ 
T5 encoder only & 11.3 \\ 
\midrule
\multicolumn{2}{c}{with Speech2Text Adapter} \\
\hdashline
Whole T5        & 9.7 \\
T5 encoder+emb  & 9.4 \\ 
T5 encoder only & 9.5 \\ 
\bottomrule
\end{tabular}
\caption{Speech recognition performance. We compare model performances with and without pretrained (see section \ref{adapter}) Speech2Text adapters, with and without retrieved entities. The WER values here were calculated from an ASR-only setup (different from joint DST-ASR setup in Table \ref{jga}).}
\label{asr}
\vspace{-.3in}
\end{table}

\section{Conclusions}
We proposed a joint speech and language model (SLM) with both speech and text inputs. The speech input is encoded using a separately trained CTC encoder where the input is down-sampled by filtering out the blank symbols from the CTC decoder. The CTC-based blank filtering reduces the number of speech frames to roughly match the textual units, unlike previous work where text was up-sampled to speech~\cite{rosenberg_etal_2019_speech}. A Speech2Text adapter, trained with seq-to-seq ASR task, transforms the blank-filtered speech encodings to the textual encoding space. This allows us to readily use pre-trained large language models for understanding the content of both speech and text inputs. The model can be trained to perform both recognition and downstream speech understanding task simultaneously. 
Our results, on both DST and ASR tasks, comparing training different groups of parameters show that the LLM decoder does not need to be trained to obtain most of the performance gains from the Speech2Text adapter. This suggests that the adapter is effective in bringing the speech to the text encoder space.

Further, we introduce an Speech2Entity retriever to select entities relevant to the speech input using a two-tower model with the SLM encoder. In our retrieval-based SLM (ReSLM), by pre-pending the retrieved entities to the text input, we show that the performance of inferring the dialog states related to task-specific entities can be improved. This also translates to significant improvement in the downstream speech understanding task, in our case, prediction of dialog states ($34.5\%$ JGA). Thus, the combined system with the Speech2Text adapter and the Speech2Text retriever outperforms a strong cascade baseline system ($31.8\%$ JGA) where the DST was trained on error-prone ASR transcripts. Similarly, the ReSLM model ($8.6\%$ WER) with the adapter and the retriever outperforms a strong in-domain ASR baseline ($10.4\%$ WER). 

While the experiments are performed on DST task, the model is more widely applicable and its performance can be further improved with better retriever.

\section{Acknowledgements}
We would like to acknowledge Jeffrey Zhao, Abhinav Rastogi and Aramys Miranda for their invaluable help.

\bibliographystyle{IEEEtran}
\bibliography{mybib}

\end{document}